\documentclass{article}
\usepackage{amsmath}
\usepackage{amsfonts}
\usepackage{graphicx}

\newcommand{\R}{\mathbb{R}}
\def\Z{{Z\kern-.5em{Z}}}
\def\V{{V\kern-.7em{V}}}

\def\d{\partial}

\def\bvec#1{{#1}}      
\def\vb{\bvec{b}}
\def\vk{\bvec{k}}
\def\vq{\bvec{q}}

\def\dk#1#2{\frac{ d^{#2}{#1} }{ (2\pi)^{#2} }} 
\def\da#1#2{\frac{ d{#1}}{{#1}^{{#2}+1}}}
\begin{document}
\title{Scale-dependent stochastic quantization}

\author{Mikhail Altaisky \thanks{Talk given at the International 
Conference ``Frontiers of Fundamental and Computational Physics, FFP6'', 
September 2004, Udine, Italy.} \\
\small Joint Institute for Nuclear Research, Dubna, 141980, Russia; and \\
\small Space Research Institute, Moscow, 117997,Russia\\ 
\small E-mail: altaisky@mx.iki.rssi.ru}
\date{Sep 28, 2004}
\maketitle

\begin{abstract}
Based on the 
wavelet-defined multiscale random noise proposed in \cite{dan}, a multiscale 
version of the stochastic quantization procedure is considered. 
A new type of the commutation relations emerging from the 
multiscale decomposition of the operator-valued fields is derived. 
\end{abstract}

\section{Introduction}
A highly original method of stochastic quantization of gauge fields 
proposed by G.Parisi and Y.Wu \cite{PW1981} have been attracting attention for
more than 20 years. 
Let $S_E[\phi]$ be the action Euclidean field theory in $\R^d$.
Then, instead of direct calculation of the Green functions  
from the generation functional of the field theory, it is possible to 
introduce a {\em fictitious time} variable $\tau$, make the quantum 
fields into stochastic fields 
$\phi(x) \to \phi(x,\tau),x\!\in\!\R^d,\tau\!\in\!\R $ and evaluate the 
moments  
$\langle \phi(x_1,\tau_1) \ldots \phi(x_m,\tau_m) \rangle$ by averaging over 
a random process $\phi(x,\tau,\cdot)$ governed by the Langevin equation 
\begin{equation}
\frac{\d\phi(x,\tau)}{\d\tau} + \frac{\delta S}{\delta\phi(x,\tau)} 
= \eta(x,\tau).
\label{le}
\end{equation}  
The Gaussian random force is 
$\delta$-correlated in both the $\R^d$ coordinate and 
the fictitious time:
\begin{equation}
\langle \eta(x,\tau)\eta(x',\tau') \rangle = 2 D_0 
\delta(x-x')\delta(\tau-\tau'),\quad \langle \eta(x,\tau)\rangle=0. 
\label{wn1}
\end{equation}
The physical Green functions are obtained by taking the steady state limit
$$
G(x_1,\ldots,x_m) = \lim_{\tau\to\infty} 
\langle \phi(x_1,\tau) \ldots \phi(x_m,\tau) \rangle.
$$

Following \cite{g24} we extend the method of stochastic quantization  
by introducing the scale-dependent random processes $W(a,b,\cdot)$, 
where $b \in \R^d$ is a spatial coordinate, and $a$ is the spatial resolution. 
For a square-integrable function $f(x,\cdot)$ the wavelet 
coefficients are 
\begin{equation}
W(a,\vb,\cdot) = \int |a|^{-d}
\overline{\psi\left(\frac{x-b}{a}\right)}f(x,\cdot)d^dx.
\label{wtr1}
\end{equation}
Hereafter they will be referred to as the {\em scale components} of $f$ with 
respect to the basic wavelet $\psi$. The reconstruction of a  
function from its scale components is given by the inverse wavelet 
transform 
\begin{equation}
f(x,\cdot) = \frac{2}{C_\psi} \int_0^\infty \frac{da}{a^{d+1}}\int d^db  
\psi\left( \frac{x-b}{a}\right) W(a,b,\cdot), 
\label{fa1}
\quad 
C_\psi = \int \frac{|\tilde\psi(k)|^2}{S_d|k|^d}d^dk,
\end{equation}
with $S_d$ being the are of the unit sphere in $d$ dimensions, 
is the normalization for the isotropic wavelets. 
Performing the wavelet transform (in spatial coordinate) of the 
fields and the random force in the Langevin 
equation, we get the possibility to substitute the white noise \eqref{wn1}
by a scale-dependent random force
\begin{equation}
\langle \tilde \eta(a_1,k_1,\tau_1) \tilde \eta(a_2,k_2,\tau_2) \rangle = 
C_\psi (2\pi)^d \delta^d(k_1+k_2) \delta(\tau_1-\tau_2) a_1 
\delta(a_1-a_2) D(a_1,k_1)
\label{gns1}
\end{equation}
In case the spectral density of the random force  is a constant 
$D(a_1,k_1)=D_0$, the inverse wavelet transform 
\begin{equation}
\phi(x)  = \frac{2}{C_\psi} \int_0^\infty \frac{da}{a} \int \dk{k}{d}
\frac{d\omega}{2\pi} 
\exp(\imath(kx-\omega\tau)) 
\tilde \psi(ak) \tilde \phi(a,k,\omega),
\label{zt}
\end{equation}
drives the process \eqref{gns1} into the white noise \eqref{wn1}. 

In case of arbitrary functions $\phi(a,x,\cdot)$ we have 
more possibilities. In particular, 
we can define a narrow band forcing that acts at a single scale 
\begin{equation}
D(a,k) = a_0 \delta(a-a_0) D_0.
\label{sb}
\end{equation} 
The contribution of the scales with the wave vectors apart from 
the the typical scale $a_0^{-1}$ is  suppressed by rapidly vanishing 
wings of the compactly supported wavelet $\tilde\psi(k)$. 

Here we present two examples of the divergence free stochastic 
perturbation expansion: (i) the scalar field theory $\phi^3$, (ii) 
the non-Abelian gauge field  theory. 

\section{Scalar field theory}
\begin{figure}[h]
\centering\includegraphics[width=8cm]{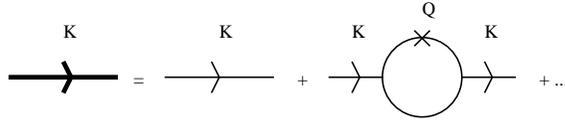}
\caption{Diagram expansion of the stochastic Green function in $\phi^3$-model}
\label{gf:pic} 
\end{figure}

Let us turn to the stochastic quantization of the $\phi^3$ theory with 
the  scale-dependent noise \cite{g24}.
The Euclidean action of the $\phi^3$ theory is 
\begin{equation}
S_E[\phi(x)]= \int d^d x \left[\frac{1}{2} (\d\phi)^2 + \frac{m^2}{2}\phi^2
+ \frac{\lambda}{3!}\phi^3 \right].
\label{pf3}
\end{equation}
The corresponding Langevin equation is written as  
\begin{equation}
\frac{\d\phi(x,\tau)}{\d\tau} + \left( 
-\Delta \phi + m^2\phi + \frac{\lambda}{2!} \phi^2 \right) = \eta(x,\tau).
\label{le3}
\end{equation} 
Substituting the scale components in representation \eqref{zt} we 
get the integral equation for the stochastic fields 
\begin{equation}
\begin{array}{l}
(-\imath\omega + k^2 + m^2)\phi(a,k,\omega) =  \eta(a,k,\omega)
-\frac{\lambda}{2}  \overline{\tilde\psi(ak)} 
\left(\frac{2}{C_\psi}\right)^2 
\int\dk{k_1}{d}\frac{d\omega_1}{2\pi}\da{a_1}{d}\da{a_2}{d}\\ 
\tilde\psi(a_1k_1) \tilde\psi(a_2(k-k_1))  
\phi(a_1,k_1,\omega_1) \phi(a_2,k-k_1,\omega-\omega_1) 
.
\end{array}
\label{phi3w}
\end{equation}
Starting from the zero-th order approximation 
$\phi_0 = G_0 \eta$ with the bare Green function  
$G_0(k,\omega) = 1/(-\imath\omega + k^2 + m^2)$
and iterating the integral equation \eqref{phi3w}, 
we get the one-loop correction to the stochastic Green function 
\begin{equation}
G(k,\omega) = G_0(k,\omega) + \lambda^2 G_0^2(k,\omega) \int 
\dk{q}{d}\frac{d\Omega}{2\pi}  2\Delta(q) 
|G_0(q,\Omega)|^2 G_0(k-q,\omega-\Omega) +\ldots,
\label{phi3G1}
\end{equation}
where $\Delta(k)$ is the scale averaged effective force correlator 
\begin{equation}
\Delta(k) \equiv  \frac{2}{C_\psi} 
\int_0^\infty \frac{da}{a} |\hat\psi(ak)|^2 D(a,k).
\label{dak}
\end{equation}  
In the same way the other stochastic momenta can be evaluated. Thus the 
common stochastic 
diagram technique is reproduced with the scale-dependent random force
\eqref{gns1} instead of the standard one \eqref{wn1}.
The 1PI diagrams corresponding to the stochastic Green function decomposition 
\eqref{phi3G1} are shown in Fig.~\ref{gf:pic}.

It can be easily seen that for a single-band forcing \eqref{sb} and a suitably 
chosen wavelet the loop divergences are suppressed. For instance, the 
use of the Mexican hat wavelet  
\begin{equation}
\hat \psi(k) = (2\pi)^{d/2} (-\imath k)^2 \exp(-k^2/2), \quad 
C_\psi = (2\pi)^d
\label{mh}
\end{equation}
for the single band random force \eqref{sb}
gives the effective force correlator 
\begin{equation}
\Delta(\vq) = (a_0q)^4 e^{-(a_0q)^2}D_0
\label{efc2}.
\end{equation}
The loop integrals taken with this effective force 
correlator \eqref{efc2} can be easily seen to be free of 
ultra-violet divergences 
\begin{eqnarray}
\nonumber G_2(k,\omega)  &=&  G_0^2(k,\omega) \int \dk{q}{d} 2\Delta(q) 
\int_{-\infty}^\infty 
\frac{d\Omega}{2\pi} \frac{1}{\Omega^2 + (q^2+m^2)^2} \\
&\times& \frac{1}{-\imath(\omega-\Omega) + (k-q)^2+m^2}
\label{p32} 
\end{eqnarray}

\section{Non-Abelian gauge theory}
The Euclidean action of a non-Abelian field is given by 
\begin{equation}
S[A] = \frac{1}{4} \int d^d x F_{\mu\nu}^a(x)F_{\mu\nu}^a(x), 
\label{ym1} \quad
F_{\mu\nu}^a(x) = \d_\mu A_\nu^a(x) -\d_\nu A_\mu^a(x)
+g f^{abc}A_\mu^b(x) A_\nu^c(x).   
\end{equation}
The Langevin equation for the gauge theory \eqref{ym1} can be written as  
\begin{equation}
\frac{\d A_\mu^a(x,\tau)}{\d\tau} + 
\bigl( -\delta_{\mu\nu}\d^2 + \d_\mu \d_\nu
\bigr)A_\nu^a(x,\tau)  = \eta_\mu^a(x,\tau) + U_\mu^a(x,\tau),
\label{le2}
\end{equation}
where $\eta_\mu^a(x,\tau)$ is the random force and $U_\mu^a(x,\tau)$ 
is the nonlinear interaction term 
$$
U[A] = \frac{g}{2}V^0(A,A) + \frac{g^2}{6}W^0(A,A,A).
$$
The stochastic diagram technique for the gauge field Langevin equation 
\eqref{le2} is summarized in the Table~\ref{tab1}.
\begin{table}
\caption{Stochastic diagrams for the non-Abelian gauge fields. 
Redrawn from \cite{namiki}}
\begin{tabular}{|c|l|l|}
\hline
Diagram & Notation & Formula \\
\hline 
\includegraphics[width=2cm,height=0.5cm]{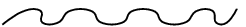}        & $G_{\mu\nu}^{ab}(k,\tau-\tau')$ &
$\delta^{ab}\theta(\tau-\tau') \left[
\left( \delta_{\mu\nu} -\frac{k_\mu k_\nu}{k^2} \right) 
e^{-k^2(\tau-\tau')}
+ \frac{k_\mu k_\nu}{k^2} 
\right]$ \\
\hline 
\includegraphics[width=2cm,height=0.8cm]{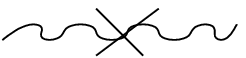}        &    $D_{\mu\nu}^{ab}(k,\tau-\tau')$ &
             $\delta^{ab}\Bigl[
             \left(\delta_{\mu\nu} -\frac{k_\mu k_\nu}{k^2} \right)
             \bigl( e^{-k^2|\tau-\tau'|} - e^{-k^2(\tau+\tau')}\bigr)$\\
        &  & $ 
             + 2 min(\tau,\tau') \frac{k_\mu k_\nu}{k^2}
             + \frac{k_\mu k_\nu}{k^2} \Bigr]
             $\\
\hline
\includegraphics[width=1.5cm]{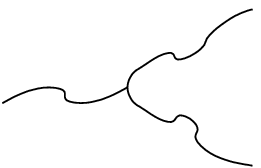}       &
        $\frac{g}{2}V_{\mu\kappa\lambda}^{abc}(k_1,k_2,k_3)$& 
$\frac{\imath g}{2}f^{abc}\bigl[
(k_1-k_2)_\lambda \delta_{\mu\kappa} + (k_2-k_3)_\mu \delta_{\kappa\lambda}$\\
       & & $
+(k_3-k_1)_\kappa \delta_{\mu\lambda}
\bigr]$ \\
\hline 
\includegraphics[width=2cm,height=15mm]{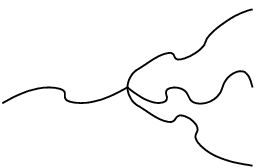}       & $\frac{g^2}{6} W_{\mu\nu\kappa\lambda}^{abcd}$ &
$-\frac{g^2}{6}\bigl[ 
f^{xab} f^{xcd}(\delta_{\mu\kappa}\delta_{\nu\lambda}- \delta_{\mu\lambda}
\delta_{\nu\kappa})$\\
    & & $
+ f^{xac} f^{xbd}(\delta_{\mu\nu}\delta_{\kappa\lambda} - \delta_{\mu\lambda}\delta_{\nu\kappa}) $\\
    & & $
+ f^{xad} f^{xbc}(\delta_{\mu\nu}\delta_{\kappa\lambda} - \delta_{\mu\kappa}\delta_{\nu\lambda}) 
\bigr]$\\
\hline
\end{tabular}
\label{tab1}
\end{table}
The two terms standing in the free field Green function correspond to 
the transversal and the longtitudal mode propagation:
$$
G_{\mu\nu}^{ab}(k) = \frac{T_{\mu\nu}(k)\delta_{ab}}{-\imath\omega + k^2} 
                   + \frac{L_{\mu\nu}(k)\delta_{ab}}{-\imath\omega}, \qquad
T_{\mu\nu}(k) = \delta_{\mu\nu} - \frac{k_\mu k_\nu}{k^2}, L_{\mu\nu}(k)=\frac{k_\mu k_\nu}{k^2}. 
$$
(Here we are concerned with divergences and do not touch 
any gauge fixing.)

Similarly the scalar field theory, we can use the scale-dependent 
forcing \eqref{gns3} in the Langevin equation \eqref{le2}. 
Since there is no dynamic evolution for the longtitudal modes 
in the Langevin equation \eqref{le2}, it is natural to use the 
transversal scale-dependent random force 
\begin{eqnarray}
\nonumber \langle \eta_\mu^a (a_1,k_1,\tau_1) \eta_\nu^b(a_2,k_2,\tau_2) 
\rangle &=& (2\pi)^d \delta^d(k_1+k_2) \delta(\tau_1-\tau_2) 
T_{\mu\nu}(k_1)\\  
&\times& C_\psi a_1 \delta(a_1-a_2) D(a_1,k_1).
\label{gns3}
\end{eqnarray}
Let us consider a gluon loop with two cubic vertices. Summing up 
over the gauge group indices 
$ \bigl(\frac{\imath}{2}g\bigr)^2 f^{abc}\delta_{bd}f^{der}\delta_{cr}=
\frac{g^2}{4}\delta_{ae}C_2,$
with $C_2=N$ for $SU_N$ groups, we can wright the gluon loop as a sum 
of two diagrams -- those with the transversal and the longtitudal 
stochastic Green functions 
\begin{equation}
G_{2\mu\nu}^{ab}(k,\omega)= g^2\delta_{ab}C_2 |G_0(k,\omega)|^2
\sum_{I=T,L}  \int \frac{d\Omega}{2\pi} \dk{q}{d} 
N^I(k,\omega,q,\Omega)l^I_{\mu\nu}(k,q) 2 \Delta(q)
\end{equation}  
where
\begin{eqnarray*}
N(k,q) = \left|\frac{1}{-\imath\Omega+q^2}\right|^2 
\begin{pmatrix}
\frac{1}{-\imath(\omega-\Omega)+(k-q)^2} \cr
\frac{1}{-\imath(\omega-\Omega)} 
\end{pmatrix}
\\
l_{\mu\nu}(k,q) = V_{\mu\kappa\lambda}(k,k-q,q) T_{\lambda\gamma}(q)
V_{\sigma\nu\gamma}(k-q,k,-q)
\begin{pmatrix}
T_{\kappa\sigma}(k-q) \cr 
L_{\kappa\sigma}(k-q)
\end{pmatrix}
\end{eqnarray*}
As it can be observed after explicit evaluation of the tensor 
structures $l^T_{\mu\nu}$ and $l^L_{\mu\nu}$,
and integration over $d\Omega$, the  wavelet factor in 
the effective force correlator $\Delta(q)$ will suppress the divergences 
for a narrow-band forcing \eqref{sb}. The power factor $k^n$ of the basic 
wavelet $\psi$, that provides $\tilde\psi(0)=0$, also makes the IR behavior 
softer. In this respect the wavelet regularization is different from 
the continuous regularization  $\int d^dy R_\Lambda(\d^2) \eta(y,\tau)$, see 
e.g. \cite{h93},  that makes UV behavior softer by the factor 
$e^{-\frac{k^2}{\Lambda^2}}$, but do not affect the IR behavior. 

\section{Commutation relation}
The stochastic quantization with a forcing localized at 
a given scale $a_0$ is in some way similar to the
lattice regularization with the mesh size of order $a_0$. However there is 
a question  what is the physical sense of the scale components, and what 
are the implications for canonical quantization of these  
fields? The answer to the first question stems from the definition of 
wavelet transform: the scale component $\phi(a,x)$ is a projection 
of the state vector $\phi$ to a certain multiresolution space \cite{mallat}, 
where $\psi$ is a basis, i.e., the basic wavelet stands for the 
aperture of the microscope by which we perceive the system $\phi$.  
To clarify the second question one can use the wavelet decomposition 
\begin{eqnarray}
\nonumber \phi(x) = \frac{2}{C_\psi} \int_0^\infty \frac{da}{a}
\int_{k_0>0}\dk{k}{d} \tilde\psi(ak) 
\bigl[ 
 e^{\imath k x} u(a,k) + (-1)^d e^{-\imath k x} u(a,-k)
\bigr], \label{wlop} \\
u^+(a,k) \equiv u(a,k)|_{k_0>0}, \quad u^-(a,k) \equiv u(a,-k)|_{k_0>0},
\label{cran}
\end{eqnarray}
where the positive and the negative energy components \eqref{cran} are 
summed up into the known plane wave components 
$$u^\pm (k) = \frac{2}{C_\psi} \int_0^\infty \frac{da}{a} \tilde\psi(ak) 
u^\pm(a,k).$$
The canonical quantization of a scalar massless field, 
the implies the  commutation relations 
\begin{equation}
[u^+(k_1),u^-(k_2)] = (2\pi)^d \delta(k_1-k_2),
\end{equation}
that can be maintained if we set \cite{ffp5} 
\begin{equation}
[u^+(a_1,k_1),u^-(a_2,k_2)] = (2\pi)^d \delta(k_1-k_2) \frac{C_\psi}{2} 
                               a_1 \delta(a_1-a_2).
\label{cr0}
\end{equation}
For a massive field, with the given energy of the free particle 
$\omega_k = \sqrt{\vk^2+m^2}$, 
the commutation relations for creation and annihilation operators 
\begin{equation}
[b(a_1,k_1),b^+(a_2,k_2)] = (2\pi)^{d-1} \omega_k\delta^{d-1}(k_1-k_2) 
C_\psi a_1 \delta(a_1-a_2).
\label{crm}
\end{equation}
To keep the Lorentz invariance at all scales the basic wavelet 
$\psi$ can depend only on Lorentz scalars, such as 
$k_\mu k^\mu = m^2$. Being compactly 
supported in both $x$ and $k$ spaces the wavelet filter 
$\tilde \psi(ak) \equiv \mu(a^2m^2)$ suppresses the contribution of the 
scale components which are far  from the 
typical scale $a_m = m^{-1}$.

It should be emphasized that the commutation relations for scale components 
(\ref{cr0},\ref{crm}) are not unique: there may be constructed some other 
commutation relations is wavelet space that maintain the same 
canonical commutation relations in wavenumber space.

As it concerns the causality and operator ordering, the introduction 
of the scale argument in operator-valued functions implies the 
operators should be ordered in both the time and the scale. 
Extending the causality in this way it was suggested \cite{ffp5} to arrange 
the operator products by decreasing scale from right to left; so that the 
rightmost operator should correspond to the largest outermost object
\begin{equation}
T( A(\Delta_x,x) B(\Delta_y,y) ) = \begin{cases}
A(\Delta_x,x) B(\Delta_y,y) & y_0 < x_0 \\
\pm B(\Delta_y,y) A(\Delta_x,x) & x_0 < y_0 \\
A(\Delta_x,x) B(\Delta_y,y) & \Delta_y > \Delta_x, y_0 = x_0 \\
\pm B(\Delta_y,y) A(\Delta_x,x) & \Delta_y < \Delta_x,x_0 = y_0. 
\end{cases}
\end{equation}


\begin{thebibliography}{9}
\bibitem{PW1981} G.Parisi and Y.-S.Wu. \emph{Scientica Sinica}, v.24, p.483 (1981)
\bibitem{dan} M.V.Altaisky. \emph{Doklady Physics}, v.48, p. 478 (2003)
\bibitem{g24} M.V.Altaisky. in \emph{Group 24: Physical and mathematical 
              aspects of symmetries} pp.893-897, IoP, (2002)
\bibitem{namiki} M. Namiki. \emph{Stochastic quantization}, Springer (1992)
\bibitem{ffp5}   M.V.Altaisky. \emph{Renormalization group and geometry}. 
                 Proc. Int. Conf. FFP5, Hyderabad (2003)
\bibitem{h93}    M.B.Halpern. \emph{Progr. Theor. Phys. Suppl.} v.111, p.163 
                 (1993) 
\bibitem{mallat} S. Mallat, \emph{A theory for multiresolution signal
		      decomposition: wavelet transform}. 
                      Preprint GRASP Lab. Dept. of Computer an
		      Information Science, Univ. of Pensilvania. (1986)
\end{thebibliography}
\end{document}